\documentstyle[preprint,aps]{revtex}
\begin{document}
\renewcommand{\thesection}{\Roman{section}}
\renewcommand{\thesubsection}{\Alph{subsection}}
\setlength{\baselineskip}{0.5cm}
\begin{center}
 {\Large \bf Low Energy Effective Action of Lightly Doped Two-Leg t-J Ladders}
\vskip 1cm
 {\bf Yu-Wen Lee\footnote{Email address: d857312@phys.nthu.edu.tw} and
  Yu-Li Lee\footnote{Email address: yllee@phys.nthu.edu.tw}}
\vskip 0.5cm
 {Physics Department, National Tsing Hua University, Hsinchu, Taiwan}
\end{center}
\vskip 1cm
\begin{abstract}
 We propose a low energy effective theory of lightly doped two-leg t-J ladders 
with the help of slave fermion technique. The continuum limit of this model consists 
of two kinds of Dirac fermions which are coupled to the $O(3)$ non-linear sigma 
model in terms of the gauge coupling with opposite sign of "charges". In addition to
the gauge interaction, there is another kind of attractive force between these Dirac 
fermions, which arises from the short-ranged antiferromagnetic order. We show 
that the latter is essential to determine the low energy properties of lightly doped 
two-leg t-J ladders. The effective Hamiltonian we obtain is a bosonic Gaussian 
model and the boson field basically describes the particle density fluctuation. We also 
find two types of gapped spin excitations. Finally, we discuss the possible 
instabilities: charge density wave (CDW) and singlet superconductivity (SC). We 
find that the SC instability dominates in our approximation. Our results indicate that 
lightly doped ladders fall into the universality class of Luther-Emery model.
\end{abstract}

71.27.+a, 71.10.Fd, 71.10.Pm

\newpage

\section{Introduction}
\vskip 0.5cm

  The properties of ladder systems described by t-J and Hubbard models have been
the subject of intensive studies recently\cite
{Dago1,Dago2,Rice1,Sigr,Noac,Khve,Hayw,Poil,Troy,Rice2,Mull,Ichi,Ivan}. The 
reason is that there are systems such as $(VO)_{2}P_{2}O_{7}$\cite{John} and 
$SrCu_{2}O_{3}$\cite{Hiro} which can be the possible realization of these 
models. Experiments on magnetic susceptibility and neutron inelastic scattering 
show the existence of a finite spin gap. Moreover, a recent measurement shows the 
sign of superconductivity in the doped spin ladders\cite{Ueha}. Therefore, the study 
of these systems can offer new insights into the nature of magnetic, charge density, 
and pairing correlations in strongly correlated electron systems.
 
 The undoped ladder systems show unusual magnetic behaviors. When the number 
of chains is even, there is a spin gap. When it is odd, the spin excitation is gapless. 
This can be easily understood by considering the limit of strong rung interactions. 
For two-leg ladders, since the spins on every rung forms a singlet first in this limit, 
it can be considered as a set of weakly coupled rung singlets. The spin excitation is 
formed by turning over one spin on a rung and this costs finite energy. Along the 
same reasoning, a three-leg ladder is effectively equivalent to a spin-$1/2$ chain and 
the latter has gapless spin excitations. Numerical studies\cite{Barn} show that the 
spin gap for two-leg ladders persists even at the experimentally interested isotropic 
point. (By this we mean that the rung interaction is equal to the intrachain interaction.) 
This indicates that even at the isotropic point, the rough picture of a ground state 
dominated by rung singlets is robust. People have investigated spin ladders by the 
semiclassical (large spin) approach\cite{Sier}. Remarkably, this approach qualitatively 
captures the basic feature of this system as in the case of spin chains. Besides, the 
spectrum predicted by the non-linear sigma model is in good agreement with other 
approximations\cite{Reig} both in the strong and weak rung interaction limit. These 
facts imply that the non-linear sigma model correctly describes the low energy sector 
of the spin liquid state of ladder systems though it ignores fluctuations along the rung. 
This is one of the motivations that we would like to use the large spin approach to 
study the lightly doped case.
 
 The effect of doping on the antiferromagnetism is an important and unsettled issue in 
strongly correlated electron systems. The main difficulty lies in hole motions, which 
cause frustration of the antiferromagnetic (AF) order. Theoretically, we have no 
adequate analytical methods to deal with the competition between charge and magnetic 
fluctuations because there is no obvious small parameter which facilitates an expansion 
about a tractable model. In the present paper, we employ the large spin expansion 
which permits us to map the original model to a continuum field theory. And this allows 
us to tackle the problem with analytical methods due to the 1-d nature of this system. 
Mean field studies\cite{Dago2,Rice1,Sigr} and numerical calculations\cite
{Hayw,Poil,Troy} show that the spin gap in undoped two-leg ladders still persists at 
low doping concentration. This implies that the underlying short-ranged AF order is not 
destroyed too much by hole motions as the doping concentration is low. Thus, after we 
resolve the constraint in the slave fermion representation, we assume that spin variables 
have a strong short-ranged AF order. This leads to the $t^{'}-J$ model proposed by 
Wiegman, Wen, Lee, and Shankar\cite{WWSL,Shan}. Because of this background AF 
order, it is natural to say that there are two kinds of holes (on A and B sublattice) with 
opposite sign of (fictitious) charges and they are coupled to the staggered magnetization 
through the Berry phase. In contrast to previous studies\cite{WWSL}, we keep the 
nearest-neighbor attractive four-fermion interactions between A and B holes. This 
attractive force can be understood as the following: the energy of two holes on the same 
spin singlet is lower than that of two holes on different singlets because there are less
broken bonds in the former. And this is equivalent to an attractive force between holes 
on different sublattices. We discuss this model in the absence of the quartic fermion 
interaction first. To study the low energy physics we can linearize the dispersion relation 
of fermions about the fermi points . It turns out that we have four branches of massless 
Dirac fermions coupled to the non-linear sigma model. From the study of Schwinger 
model, we know that massless Dirac fermions will screen the long range Coulomb force 
or give a mass term to the gauge field on account of the chiral anomaly. Therefore, 
the gauge fields are in the Higgs phase\cite{Higg}. Because of the Gauss's law, all 
excitations must be gauge singlets (or mesons in the jargon of particle physics). In the 
present case, we have three gapless spin-singlet collective modes. There are also gapped
spin-$1/2$ excitations which carry the electronic charge but no magnons. The SC 
instability is enhanced. Then we switch on the nearest-neighbor attractive four-fermion 
interactions. Two of the gapless modes become massive and the long range Coulomb 
force is not screened, i.e. the gauge fields are in the confining phase. Consequently, we 
have only one gapless charge mode, which is spin-singlet. We also have gapped 
spin-triplet excitations and electron-like collective modes, which are spin-$1/2$ and 
carry electronic charges. We compute the exponents of pairing correlation function and 
$4k_{F}$ CDW susceptibility. In our approximation, turning on the quartic fermion 
interactions enhances the tendency toward SC relative to the tendency toward CDW. 
Therefore, a weak interladder interaction will lead to SC, which has been predicted by 
other approaches\cite{Dago2,Rice1,Sigr,Hayw,Poil,Troy}. Our analysis indicates that 
these two types of attractive forces play different roles in two-leg t-J ladders. While the 
formation of spin gap and spin-hole bound states are due to the gauge interaction, 
four-fermion interactions between A and B holes are responsible for the pairing between 
holes. In addition, inclusion of the latter drastically changes the low energy properties of 
doped ladders such that they fall into the universality class of Luther-Emery 
model\cite{LE}. Our results confirm the conclusions from numerical studies\cite{Troy}.

 The rest of the paper is organized as follows. In section two we derive the low
energy effective action. In section three we discuss the implications of this 
action. We study the effect of gauge interaction in subsection A. In subsection
B we take into account the four-fermion interaction between A and B holes.
Section four is the conclusion. We give a summary of the bosonization rules we used in the
appendix.

\section{Derivation of the Effective Action}
\vskip 0.5cm

  Since the first part of our derivation is valid for general t-J models, we 
will not write down the ladder index explicitly until it is necessary. We start
from the following model
\begin{eqnarray}
 H &=& -t \ \displaystyle{\sum_{\langle i,j\rangle}}X_{\sigma 0}(i)X_{0 \sigma}
       (j)+\displaystyle{\sum_{i}}\epsilon_{d}X_{\sigma \sigma}(i)+ \nonumber \\
   & & \frac{J}{4} \ \displaystyle{\sum_{\langle i,j\rangle}}[X_{\sigma \sigma^{
       '}}(i)X_{\sigma^{'} \sigma}(j)-\frac{1}{2}X_{\sigma \sigma}(i)X_{\sigma^{
       '}\sigma^{'}}(j)]
\label{cc}
\end{eqnarray}
where $\langle i,j \rangle$ means the nearest neighbor sites, $X_{ab}\equiv 
\mid a \rangle \langle b \mid$ and $\mid a \rangle = \mid 0 \rangle$, 
$\mid \uparrow \rangle$, $\mid \downarrow \rangle$ corresponding to the empty 
site and spin-up (spin-down) sites. Since transitions between empty and occupied 
states include a change in the fermonic number, the operators $X_{\sigma 0}(i)$, 
$X_{0 \sigma}(i)$ are fermonic and the operators $X_{\sigma \sigma^{'}}(i)$, 
$X_{00}(i)$ are bosonic. It is easy to check that they satisfy the following graded 
Lie algebra called $Spl(1,2)$\cite{Tsve}:
\begin{equation}
 [X_{ab}(i) , X_{cd}(j)]_{\pm}=\delta_{ij}(\delta_{bc} X_{ad}(i)\pm \delta_{ad}
 X_{cb}(i))
\label{cd}
\end{equation}
where $(+)$ should be used only if both operators are fermonic. In the limit of
zero doping, $\epsilon_{d}\rightarrow -\infty$ and the model is reduced to the
spin-$1/2$ Heisenberg model.

  Among various methods to deal with eqs.(\ref{cc}) and (\ref{cd}), there are two 
popular representations of the above graded Lie algebra - - slave fermion and slave 
boson. We adopt the former and introduce a vacuum state annihilated by operators 
$f_{i}$ and $b_{\sigma}(i)$. Then the $X$-operators can be represented as follows 
\begin{eqnarray}
 X_{0 \sigma}(i) &=& f^{+}_{i}b_{\sigma}(i), \nonumber \\
 X_{\sigma 0}(i) &=& f_{i}b^{+}_{\sigma}(i), \nonumber \\
 X_{\sigma \sigma^{'}}(i) &=& b^{+}_{\sigma}(i)b_{\sigma^{'}}(i), \nonumber \\
 X_{00}(i) &=& f^{+}_{i}f_{i}
\label{aa}
\end{eqnarray}
with the constraint $b^{+}_{\sigma}(i)b_{\sigma}(i)+f^{+}_{i}f_{i}=1$ and $\sigma=
1,-1$ for spin up and down, respectively. Here $f_{i}$, $f^{+}_{i}$ satisfy canonical 
anticommutation relations and $b_{\sigma}(i)$, $b^{+}_{\sigma}(i)$ satisfy canonical
commutation relations. To use the large spin expansion, we replace the above constraint 
with this one: $b^{+}_{\sigma}(i)b_{\sigma}(i)+f^{+}_{i}f_{i}=2S$, which is called 
the spin-s representation of $Spl(1,2)$. Now the $X$-operators represent transitions
between states $\mid s \rangle$ and $\mid s-1/2 \rangle$. With the help of 
eq.(\ref{aa}), we can write down the path integral representation of the $t-J$ 
model 
\begin{eqnarray}
 Z &=& \int D[f]D[f^{+}]D[b_{\sigma}]D[b^{+}_{\sigma}]\delta (b^{+}_{\sigma}(i)
       b_{\sigma}(i)+f^{+}_{i}f_{i}-2S)exp(-I), \nonumber \\
 I &=& \int^{\beta}_{0}d\tau \ \displaystyle{\sum_{i}}[f^{+}_{i}(\partial_{\tau}
       -\mu)f_{i}+b^{+}_{\sigma}(i)\partial_{\tau}b_{\sigma}(i)]+\int^{\beta}_{
       0}d\tau \ H, \nonumber \\
 H &=& t \ \displaystyle{\sum_{\langle i,j \rangle}}f^{+}_{j}f_{i}b^{+}_{\sigma}
       (i)b_{\sigma}(j)+ \nonumber \\
   & & \frac{J}{4} \ \displaystyle{\sum_{\langle i,j \rangle}}[b^{+}_{\alpha}(i)
       b_{\beta}(i)b^{+}_{\beta}(j)b_{\alpha}(j)-\frac{1}{2}b^{+}_{\alpha}(i)
       b_{\alpha}(i)b^{+}_{\beta}(j)b_{\beta}(j)].
\end{eqnarray}
In the above eq., $\epsilon_{d}$ has been absorbed into $\mu$, the chemical potential of 
holes in terms of the constraint in the path integral measure. Based on the property of 
Grassmann variables: $(f^{+}f)^{2}=0$, we can resolve the constraint in the 
measure\cite{Tsve} as the following
\begin{eqnarray}
 b_{1} &=& \sqrt{2S}(1-\frac{1}{4S}f^{+}f)\cos{\frac{\theta}{2}} \exp{\frac{i}
           {2}(\chi -\phi)}, \nonumber \\
 b_{2} &=& \sqrt{2S}(1-\frac{1}{4S}f^{+}f)\sin{\frac{\theta}{2}} \exp{\frac{i}
           {2}(\chi +\phi)}
\end{eqnarray}
where ${\bf S}=(S-\frac{1}{2}f^{+}f){\bf \Omega}$ and ${\bf \Omega}=(\sin{\theta}
\cos{\phi},\sin{\theta}\sin{\phi},\cos{\theta})$. With the new variable ${\bf \Omega}$, we 
can rewrite the partition function as follows
\begin{eqnarray}
 Z &=& \int D[{\bf \Omega}]D[f]D[f^{+}]exp(-I), \nonumber \\
 I &=& \int^{\beta}_{0}d\tau \ \displaystyle{\sum_{j}}[f^{+}_{j}(\partial_{\tau}
       -\mu)f_{j}+i(S-\frac{1}{2}f^{+}_{j}f_{j}){\bf A}({\bf \Omega}_{j})
       \cdot \partial_{\tau}{\bf \Omega}_{j}] \nonumber \\
   & & +\int^{\beta}_{0}d\tau \ H, \nonumber \\
 H &=& \overline{t} \ \displaystyle{\sum_{\langle i,j \rangle}}f^{+}_{j}f_{i}
       \langle {\bf \Omega}_{i} \mid {\bf \Omega}_{j} \rangle_{1/2} + \nonumber \\
   & & \frac{\overline{J}}{8} \ \displaystyle{\sum_{\langle i,j \rangle}}(1-
       \frac{1}{2S}f^{+}_{i}f_{i})(1-\frac{1}{2S}f^{+}_{j}f_{j}){\bf \Omega}_{i}\cdot 
       {\bf \Omega}_{j}
\label{ab}
\end{eqnarray}
where $\overline{t}\equiv 2St$ and $\overline{J}\equiv 4JS^{2}$. (Note that when
$S=1/2$, $\overline{t}=t$ and $\overline{J}=J$.) $\langle {\bf \Omega}_{1} \mid 
{\bf \Omega}_{2} \rangle_{S}$ represents the overlap of two spin-$S$ coherent states, 
$\mid {\bf \Omega}_{1} \rangle$ and $\mid {\bf \Omega}_{2} \rangle$. ${\bf A}
({\bf \Omega})$ is the monopole vector potential, which satisfies $\bigtriangledown 
\times {\bf A}={\bf \Omega}$. (Here we have chosen the gauge such that
${\bf A}\cdot \partial_{\tau}{\bf \Omega}=-\cos{\theta}\partial_{\tau}\phi$
and $\chi$ is independent of $\tau$.)

  Now we focus ourselves on two-leg $t-J$ ladders. Using the notation of 
eq.(\ref{ab}), the action can be written as follows
\begin{eqnarray}
 I &=& \int^{\beta}_{0}d\tau \ \displaystyle{\sum_{m}\sum_{j}}[f^{+}_{j,m}
       (\partial_{\tau}-\mu)f_{j,m}+i(S-\frac{1}{2}f^{+}_{j,m}f_{j,m}) \nonumber
       \\
   & & {\bf A}({\bf \Omega}_{j,m})\cdot \partial_{\tau}{\bf \Omega}_{j,m}]+\int^{
       \beta}_{0}d\tau \ H, \nonumber \\
 H &=& \overline{t} \ \displaystyle{\sum_{m}\sum_{\langle i,j \rangle}}
       f^{+}_{j,m}f_{i,m}\langle {\bf \Omega}_{i,m}\mid {\bf \Omega}_{j,m}\rangle_{1/2}
       +\overline{t}_{\bot} \ \displaystyle{\sum_{j}}[f^{+}_{j,2}f_{j,1}\langle 
       {\bf \Omega}_{j,1}\mid {\bf \Omega}_{j,2}\rangle_{1/2}+h.c.] \nonumber \\
   & & +\frac{\overline{J}}{4} \ \displaystyle{\sum_{m}\sum_{j}}(1-\frac{1}{2S}
       f^{+}_{j,m}f_{j,m})(1-\frac{1}{2S}f^{+}_{j+1,m}f_{j+1,m}){\bf \Omega}_{j,m}
       \cdot {\bf \Omega}_{j+1,m} \nonumber \\
   & & +\frac{\overline{J}_{\bot}}{4} \ \displaystyle{\sum_{j}}(1-\frac{1}{2S}
       f^{+}_{j,1}f_{j,1})(1-\frac{1}{2S}f^{+}_{j,2}f_{j,2}){\bf \Omega}_{j,1}\cdot 
       {\bf \Omega}_{j,2}
\label{ac}
\end{eqnarray}
where $m=1,2$ is the chain index. To proceed, we have to make some assumptions. 
First of all, we assume that there is a short-ranged AF order such that we can 
parameterize ${\bf \Omega}_{j,m}$ as the following\cite{Sier}
\begin{equation}
 {\bf \Omega}_{j,m}=(-1)^{j+m}\sqrt{1-a^{2}{\bf L}^{2}_{m}}
                   {\bf n}(x)+a{\bf L}_{m}(x)
\label{ad}
\end{equation}
where ${\bf n}^{2}=1$, ${\bf n}\cdot {\bf L}_{m}=0$ and $a$ is the lattice 
spacing. ${\bf n}$ is the order parameter and ${\bf L}_{m}$ are the fast modes 
which will be integrated out. Comparing the $t$-term and $J$-term in eq.(\ref{ac}),
it is clear that in the large spin limit $J$-term dominates and thus eq.(\ref{ad}) is
valid. For small spins, we expect it is still a good assumption if there are strong short
range AF correlations and the latter is reflected by the existence of a finite spin gap 
in two-leg ladders. However, some conditions are required for a non-vanishing spin 
gap. Firstly, according to numerical studies\cite{Hayw,Poil,Troy}, the hole 
concentration must be low. Secondly, $J/t$ should be of order one. As has been 
noticed in ref.\cite{Troy}, the Nogaoka theorem is applicable in ladder systems. This 
theorem says that the one-hole ground state is ferromagnetic at $J=0$. This phase may 
be stable when $J/t$ is very small\cite{Mull}. Besides, numerical 
studies\cite{Hayw,Poil,Troy} show that the phase separation occurs when $J/t>2$. 
Therefore, the qualitative feature of the following results may be valid for the spin 
one-half case when the ratio of $J/t$ is order of one and the hole concentration is 
low. 

   Because of the background AF order, the original lattice is divided into two 
sublattices and it is natural to have two kinds of holes. We will call them A-holes 
and B-holes when $j+m$ is even and odd, respectively. An immediate consequence 
of eq.(\ref{ad}) is that the hole cannot hop coherently between different sublattices, 
i.e. the intersublattice hopping is forbidden. This can be understood as the following:
The amplitude for a hole hopping from site ${\it i}$ to ${\it j}$ is the product of the 
overlap of {\it spin} and {\it orbital} wavefunctions, i.e. $-t_{ij}\langle {\bf \Omega}_
{j} \mid {\bf \Omega}_{i} \rangle$. In this case, it is $-t \langle {\bf \Omega}_{A}
\mid {\bf \Omega}_{B} \rangle$. In view of eq.(\ref{ad}), this is zero if we have 
perfect N\'eel order (recall that $\langle {\bf \Omega} \mid -{\bf \Omega} \rangle =0$)
and exponentially small in a system with strong short-range AF order. Thus, $t$- and 
$t_{\bot}$-terms are effectively removed from the low energy effective Hamiltonian 
in the large $S$ analysis. One of their effects is to renormalize the parameters in low 
energy physics. It seems that the hole can hop coherently within the same sublattices 
under the hypothesis eq.(\ref{ad}). However, the question of whether the hole can 
have coherent hopping is related to whether we have well-defined quasiparticles in low 
energy. As pointed out in ref.\cite{Kane}, the latter depends on the density of states of 
spin waves in low energy. It turns out that there can be no well-defined quasiparticles 
in one dimension if the spin excitation is gapless. Fortunately, there is a spin gap in 
our system. Following the calculations in ref.\cite{Kane}, there will be a sharp peak at
the energy scale $-t$ in the spectral function of one-particle Green function and this 
implies a well-defined quasiparticle band at the bottom of the hole spectrum. 
Consequently, via the interactions with spins, the holes are able to acquire a kinetic
energy, whether there is a "bare hopping term" or not. In the following, we shall 
introduce a $t'$-term, i.e. hopping within the same sublattice, to represent the kinetic
energy of holes. According to numerical calculations\cite{Troy}, $t^{'}$ is of the same 
order as $J$.

   Substituting eq.(\ref{ad}) into eq.(\ref{ac}), adding a gauge-invariant $t'$-term to 
eq.(\ref{ac}) and expanding the action to the quadratic terms of ${\bf L}_{m}$ then 
integrating out ${\bf L}_{m}$, we get to the following effective action
\begin{eqnarray}
 I &=& I_{n}+I_{h}, \nonumber \\
 I_{n} &=& \frac{1}{2g^{2}}\int^{\beta}_{0}d\tau \int dx \ (\frac{1}{v^{2}_{s}}
           \mid\partial_{\tau}{\bf n}\mid^{2}+\mid \partial_{x}
           {\bf n}\mid^{2}), \nonumber \\
 I_{h} &=& \int^{\beta}_{0} d\tau \ \displaystyle{\sum_{m}\sum_{j}}[f^{+}_{A,m}
           (j)(\partial_{\tau}-ia_{0}-\mu)f_{A,m}(j)+ \nonumber \\
       & & f^{+}_{B,m}(j)(\partial_{\tau}+ia_{0}-\mu)f_{B,m}(j)]+\int^{\beta}_{
           0}d\tau \ H_{h} 
\label{ae}
\end{eqnarray}
and
\begin{eqnarray}
 H_{h} &=& H_{0}+H_{1}+H_{2}, \nonumber \\
 H_{0} &=& \overline{t^{'}} \ \displaystyle{\sum_{m}\sum_{j}}[f^{+}_{A,m}(j+2)
           f_{A,m}(j)\exp{(ia_{x}\Delta x)}+h.c.]+ \nonumber \\
       & & (A\rightarrow B, a_{x}\rightarrow -a_{x}), \nonumber \\
 H_{1} &=& -\frac{J}{4} \ \displaystyle{\sum_{m}\sum_{j}}[:f^{+}_{A,m}(j)f_{A,m}
           (j)::f^{+}_{B,m}(j+1)f_{B,m}(j+1):+(A\leftrightarrow B)], \nonumber 
           \\     
 H_{2} &=& -\frac{J_{\bot}}{4} \ \displaystyle{\sum_{j}}[:f^{+}_{A,1}(j)f_{A,1}
           (j)::f^{+}_{B,2}(j)f_{B,2}(j):+(A\leftrightarrow B)]
\label{af}
\end{eqnarray}
where $g^{2}=2/\overline{J}a(1-\delta /2S)^{2}$, $v_{s}=\frac{\overline{J}a}{2S}
(1-\delta /2S)\sqrt{1+\frac{\overline{J}_{\bot}}{2\overline{J}}}$, $\overline{t^
{'}}\equiv 2St^{'}$, $\Delta x=2a$, and $\delta$ is the hole concentration. In 
the derivation of above eqs., we have used a property of spin coherent states: 
$\langle {\bf \Omega}_{1}\mid {\bf \Omega}_{2}\rangle_{S}\approx \exp{-iS{\bf A}
\cdot ({\bf \Omega}_{1}-{\bf \Omega}_{2}})$ when ${\bf \Omega}_{1}\approx 
{\bf \Omega}_{2}$. The gauge fields $a_{\mu}\equiv \frac{1}{2}{\bf A}\cdot 
\partial_{\mu}{\bf n}$. It is clear that A and B holes carry opposite sign of charges 
because they are on different sublattices in which the staggered magnetizations 
have opposite directions. We propose that the low energy sector of lightly doped 
two-leg ladders can be described by eqs.(\ref{ae}) and (\ref{af}).

    We conclude this section with a brief summary of the $CP^{1}$ representation of the 
non-linear sigma model\cite{Poly}, which will be used later. The order parameter 
${\bf n}$ can be parameterized with a normalized spinor as
\begin{eqnarray}
  {\bf n}=\bar{z}{\bf \sigma}z \ , \ \bar{z}z=1. \nonumber
\end{eqnarray}
The $CP^{1}$ version of the non-linear sigma model in the Euclidean space is
\begin{eqnarray}
 Z &=& \int \ D[a_{\mu}]D[z]D[\bar{z}]D[\lambda] exp(-I) , \nonumber \\
 I &=& \frac{1}{e^{2}}\int \ d^{2}x \ [\mid \partial_{\mu}z-ia_{\mu}z\mid^{2}+
            i\lambda (\mid z \mid^{2}-1)]
\label{bb}
\end{eqnarray}
where $x_{0}=v_{s}\tau$, $e^{2}=2g^{2}v_{s}$ and $\lambda$ is the Lagrangian 
multiplier. An appropriate choice of the gauge makes $a_{\mu}$ in eq.(\ref{bb}) 
equivalent to the ones in eqs.(\ref{ae}) and (\ref{af}) and thus we can use the same 
notation to denote them. In the $CP^{N}$ model at large $N$, the $z$-quanta become 
massive, i.e. $\lambda$ acquires a non-vanishing mean value. At distances larger than 
this scale, the effective action of gauge fields will have the usual Maxwell term. 
Accordingly, the low energy excitation is a massive triplet, which can be considered as 
the bound state of $\bar{z}$ and $z$.

\section{Excitation Spectrum}
\vskip 0.5cm

  To discuss the low energy physics, we linearize the dispersion relation of 
fermions about their fermi points, which satisfy $k_{F}a=\frac{\pi}{2}
(1-\delta )$. ( We have assumed that the Hamiltonian is invariant when we
interchange the chain index and A and B holes. Besides, remember that $a$ is the
lattice spacing of the original lattice.) After doing that, $H_{0}$ becomes
\begin{equation}
 H_{0}=E_{0}+v_{0}\displaystyle{\sum_{m}}\int dx \ [:\psi^{+}_{A,m}\alpha (-i
       \partial_{x}-a_{x})\psi_{A,m}:+(A\rightarrow B, a_{x}\rightarrow -a_{x})]
\end{equation}
where $v_{0}=4\overline{t^{'}}a\sin{2k_{F}a}=4\overline{t^{'}}a\sin{\pi \delta}$ is 
the fermi velocity, $\psi$ is the two-component Dirac fermion, and 
$\alpha =\sigma_{3}$. Now we can analyze the effects of different interactions.

\subsection{The Effect of Gauge Coupling}
\vskip 0.5cm

  We set $H_{1}=H_{2}=0$ first. Then we rescale the imaginary time: $v_{0}\tau
\rightarrow \tau$ and do analytical continuation to the real time formalism. The
effective action is
\begin{eqnarray}
 I &=& I_{0}+I_{n}, \nonumber \\
 I_{0} &=& \displaystyle{\sum_{m}}\int d^{2}x \ [\bar{\psi}_{A,m}\gamma^{
           \mu}(i\partial_{\mu}-a_{\mu}-eA_{\mu})\psi_{A,m}+ \nonumber \\
       & & (A\rightarrow B, a_{\mu}\rightarrow -a_{\mu})]
\end{eqnarray}
and $I_{n}$ has the same form as the one in eq.(\ref{ae}) and $\gamma_{\mu}$ is
the Dirac $\gamma$-matrices. Here  $A_{\mu}$ are the external electromagnetic 
fields. In terms of the standard bosonization rules (see the appendix), $I_{0}$ can 
be bosonized as the following
\begin{eqnarray}
 I_{0} &=& \frac{1}{2} \ \displaystyle{\sum_{m}}\int d^{2}x \ [(\partial_{\mu}
           \phi_{A,m})^{2}+(\partial_{\mu}\phi_{B,m})^{2}]- \nonumber \\
       & & \frac{e}{\sqrt{\pi}}\displaystyle{\sum_{m}}\int d^{2}x \ A_{\mu}
           \epsilon^{\mu \nu}\partial_{\nu}(\phi_{A,m}+\phi_{B,m})+ \nonumber \\
       & & \frac{1}{\sqrt{\pi}}\displaystyle{\sum_{m}}\int d^{2}x \ a_{\mu}
           \epsilon^{\mu \nu}\partial_{\nu}(\phi_{A,m}-\phi_{B,m}).
\end{eqnarray}
We define $\phi_{\pm,m}\equiv \frac{1}{\sqrt{2}}(\phi_{A,m}\pm \phi_{B,m})$.
Then $I_{0}$ can be written as 
\begin{eqnarray}
 I_{0} &=& \frac{1}{2} \ \displaystyle{\sum_{m}}\int d^{2}x \ [(\partial_{\mu}
           \phi_{+,m})^{2}+(\partial_{\mu}\phi_{-,m})^{2}]- \nonumber \\
       & & \sqrt{\frac{2}{\pi}}e \ \displaystyle{\sum_{m}}\int d^{2}x \ A_{\mu}
           \epsilon^{\mu \nu}\partial_{\nu}\phi_{+,m}+\sqrt{\frac{2}{\pi}} \ 
           \displaystyle{\sum_{m}}\int d^{2}x \ a_{\mu}\epsilon^{\mu \nu}
           \partial_{\nu}\phi_{-,m}.
\end{eqnarray}
The above action can be further simplified if we define the following fields
\begin{eqnarray}
 \Phi_{1} &\equiv&  \frac{1}{\sqrt{2}}(\phi_{+,1}+\phi_{+,2}), \nonumber \\
 \Phi_{2} &\equiv&  \frac{1}{\sqrt{2}}(\phi_{+,1}-\phi_{+,2}), \nonumber \\
 \Phi_{3} &\equiv&  \frac{1}{\sqrt{2}}(\phi_{-,1}+\phi_{-,2}), \nonumber \\
 \Phi_{4} &\equiv&  \frac{1}{\sqrt{2}}(\phi_{-,1}-\phi_{-,2}).
\end{eqnarray}
With the help of the above canonical transformation, we obtain the following low
energy effective action
\begin{eqnarray}
 I_{0} &=& \frac{1}{2} \ \displaystyle{\sum_{\alpha =1}^{4}}\int d^{2}x \ (
           \partial_{\mu}\Phi_{\alpha})^{2}-\frac{2}{\sqrt{\pi}}e \ \int d^{2}x 
           \ A_{\mu}\epsilon^{\mu \nu}\partial_{\nu}\Phi_{1} \nonumber \\
       & & +\frac{2}{\sqrt{\pi}} \ \int d^{2}x \ \Phi_{3}\epsilon^{\mu \nu}
           \partial_{\mu}a_{\nu}.
\label{ag}
\end{eqnarray}

   What can we learn from eqs.(\ref{bb}) and (\ref{ag})? Firstly, there are three 
gapless spin-singlet excitations: $\Phi_{1}$, $\Phi_{2}$ and $\Phi_{4}$. Secondly,
to understand the coupled system of $\Phi_{3}$ and $z$, we have to investigate 
the gauge field dynamics. This can be done by integrating out the $\Phi_{3}$-field
since its action is quadratic and we obtain
\begin{eqnarray}
 \frac{1}{\pi^{2}}\int \ d^{2}x d^{2}y \ \tilde{f}(x)ln\mid x-y \mid \tilde{f}(y)
 \nonumber
\end{eqnarray}
where $\tilde{f}\equiv \epsilon^{\mu \nu}\partial_{\mu}a_{\nu}$ is the dual field
strength. If we choose the Lorentz gauge: $\partial_{\mu}a^{\mu}=0$, then the 
above eq. becomes the mass term of $a_{\mu}$. That is to say, the fluctuations of
the $\Phi_{3}$-field will screen the long range Coulomb force. It follows that there
are massive spin-$1/2$ excitations and they are neutral with respect to the 
$a_{\mu}$-fields in view of the gauge field mass. (This is easy to be seen by 
integrating Gauss's law from $-\infty$ to $+\infty$.) Since the $z$-quantum has
charge $1$, where is the compensating charge from? It must come from the hole
sector. We can understand this as the following\cite{Shan}:

   Let us integrate out all fields except $\Phi_{3}$. The action we obtain must be of 
the form $\cos{8\sqrt{\pi}n\Phi_{3}}$ where $n$ is an integer. (Of course, the 
renormalization of the kinetic term is possible.) This is because the coefficient of 
$\Phi_{3}$ in eq.(\ref{ag}) is proportional to the instanton density: 
$\frac{1}{4\pi}\epsilon^{\mu \nu}\partial_{\mu}a_{\nu}$. The action is invariant 
under the transformation: $\Phi_{3} \ \rightarrow \ \Phi_{3}+\sqrt{\pi}/4$ and the 
cosine just has the form which satisfies the requirement. If the cosine is relevant, 
this translation symmetry is spontaneously broken and the value of $\Phi_{3}$ is 
pinned at some minimum of the potential. In addition, its solitons carry exactly the 
charge needed to match the charge of the $z$-quantum. The Gauss's law demands
that at least one of the cosines in question must be relevant and thus the excitations 
corresponding to the $\Phi_{3}$ sector are massive. These solitons are nothing but
the A (or B) holes. They carry the electronic charges as well as the $U(1)$ charges.
As a consequence, we expect these massive doublets also carry the electronic 
charges, i.e. they carry the same quantum numbers as electrons. We should 
emphasize that there are no $z \bar{z}$ bound states, i.e. massive triplets, in this
model. As we shall see later, the absence or presence of this type of excitations is 
one of the distinctions between the Higgs phase and confining phase in our model.

     After identifying the spectrum, we would like to examine the pairing correlation 
function. Because of the spin gap, the pair-field of singlet SC can be defined as the 
following
\begin{equation}
 \Delta(j)\equiv f_{A,1}(j)f_{B,2}(j).
\label{ah}
\end{equation}
Since the correlation functions of $\Phi_{3}$ decay exponentially, the long 
distance behavior of the pairing correlation function can be calculated by the 
following operators: 
\begin{eqnarray}
 \Delta(j) &\sim& \exp{(-i\sqrt{\pi}\Theta_{1})}\exp{(-i\sqrt{\pi}\Phi_{2})}\exp
                  {(-i\sqrt{\pi}\Theta_{4})}+ \nonumber \\
           & &    \exp{(-i\sqrt{\pi}\Theta_{1})}\exp{(i\sqrt{\pi}\Phi_{2})}\exp
                  {(-i\sqrt{\pi}\Theta_{4})}
\end{eqnarray}
where $\Theta_{\alpha}$ is the dual field of $\Phi_{\alpha}$. The pairing 
correlation function behaves as $\langle \Delta(j)\Delta^{+}(0)\rangle \stackrel
{\mid j \mid \rightarrow \infty}{\longrightarrow}\frac{1}{\mid j \mid^{3/2}}$. 
Compared with the one of free fermions, which behaves like $\frac{1}{\mid j 
\mid^{2}}$, we see that  the pairing susceptibility is enhanced. 

  To sum up, if we consider the gauge coupling only, then the low energy effective
Hamiltonian consists of three gapless spin-singlet modes and one massive 
spin-$1/2$ modes, which carry the electronic charges. (It is possible that there are
massive spin-singlet modes. However, the gapless modes and the massive doublet
predominant the long distance behavior of correlation functions.) Compared with the 
numerical results\cite{Troy}, there are too many gapless modes but no magnons. It is 
not enough to consider the gauge interactions only. We will see in the next section 
the importance of taking into account $H_{1}$ and $H_{2}$ to get the correct low 
energy properties.

\subsection{The Role of $H_{1}$ and $H_{2}$}
\vskip 0.5cm

  Here we take $H_{1}$ and $H_{2}$ into account. The continum limit of them are 
as follows
\begin{eqnarray}
 H_{1} &=& -g_{1} \ \displaystyle{\sum_{m}}\int dx \ :\psi^{+}_{A,m}\psi_{A,m}:
           :\psi^{+}_{B,m}\psi_{B,m}:+ \nonumber \\
       & & g_{1}\cos{\pi \delta} \ \displaystyle{\sum_{m}}\int dx \ 
              [\psi^{+}_{A,L,m}\psi_{A,R,m}\psi^{+}_{B,R,m}\psi_{B,L,m}+(R
           \leftrightarrow L)], \nonumber \\
 H_{2} &=& -g_{2} \ \int dx \ [:\psi^{+}_{A,1}\psi_{A,1}::\psi^{+}_{B,2}\psi_{
           B,2}:+(A\leftrightarrow B)]- \nonumber \\
       & & g_{2} \ \int dx \ [\psi^{+}_{A,R,1}\psi_{A,L,1}\psi^{+}_{B,L,2}\psi_{
           B,R,2}+\psi^{+}_{A,L,1}\psi_{A,R,1}\psi^{+}_{B,R,2}\psi_{B,L,2}
           \nonumber \\
       & & +(A\leftrightarrow B)]
\end{eqnarray}
where $g_{1}=2Ja/v_{0}$, $g_{2}=J_{\bot}a/v_{0}$ and $\psi_{L,R}$ are 
left-handed and right-handed fermions, respectively. With the same convention 
used in the previous section, the bosonized forms of the above eqs. are as 
follows
\begin{eqnarray}
 H_{1} &=& -\frac{g_{1}}{2\pi} \ \displaystyle{\sum_{m}}\int dx \ [(\partial_{x}
           \phi_{+,m})^{2}-(\partial_{x}\phi_{-,m})^{2}]+ \nonumber \\
       & & g^{'}_{1} \ \displaystyle{\sum_{m}}\int dx \ \cos{\sqrt{8\pi}\phi_
           {-,m}}, \nonumber \\
 H_{2} &=& -\frac{g_{2}}{\pi} \ \int dx \ (\partial_{x}\phi_{+,1}\partial_{x}
           \phi_{+,2}-\partial_{x}\phi_{-,1}\partial_{x}\phi_{-,2})- \nonumber 
           \\
       & & 2g^{'}_{2} \ \int dx \ \cos{\sqrt{2\pi}(\phi_{+,1}-\phi_{+,2})}\cos
           {\sqrt{2\pi}(\phi_{-,1}+\phi_{-,2})}.
\end{eqnarray}
We can do the following canonical transformation to diagonalize the quadratic 
part of the Hamiltonian
\begin{eqnarray}
 \Phi_{1} &=& \sqrt{\frac{K_{1}}{2}}(\phi_{+,1}+\phi_{+,2}), \nonumber \\
 \Phi_{2} &=& \sqrt{\frac{K_{2}}{2}}(\phi_{+,1}-\phi_{+,2}), \nonumber \\
 \Phi_{3} &=& \sqrt{\frac{K_{3}}{2}}(\phi_{-,1}+\phi_{-,2}), \nonumber \\
 \Phi_{4} &=& \sqrt{\frac{K_{4}}{2}}(\phi_{-,1}-\phi_{-,2}).
\label{ak}
\end{eqnarray}
The parameters in the above eqs. are as follows
\begin{eqnarray}
 K_{1} &=& \sqrt{1-\frac{g_{1}+g_{2}}{\pi}}, \nonumber \\
 K_{2} &=& \sqrt{1-\frac{g_{1}-g_{2}}{\pi}}, \nonumber \\
 K_{3} &=& \sqrt{1+\frac{g_{1}+g_{2}}{\pi}}, \nonumber \\
 K_{4} &=& \sqrt{1+\frac{g_{1}-g_{2}}{\pi}}.
\label{al}
\end{eqnarray}
These relations are valid only in the weak coupling limit. In our case, this 
corresponds to the large spin limit. (Remember that $\overline{t^{'}}(=2St^{'})$ 
appears in the denominator of the definition of $g_{i}$.) We can see that 
$K_{1} < 1$ and $K_{3} > 1$ because $g_{i} >0$. (This corresponds to the 
attractive force between holes.) The effective Hamiltonian of the hole sector is 
\begin{eqnarray}
 H_{h} &=& \frac{1}{2} \ \displaystyle{\sum_{\alpha =1}^{4}}K_{\alpha} \ \int dx
           \ [(\partial_{x}\Theta_{\alpha})^{2}+(\partial_{x}\Phi_{\alpha})^{2}]
           + \nonumber \\
       & & 2g^{'}_{1} \ \int dx \ \cos{\sqrt{\frac{4\pi}{K_{3}}}\Phi_{3}}\cos
           {\sqrt{\frac{4\pi}{K_{4}}}\Phi_{4}}- \nonumber \\
       & & 2g^{'}_{2} \ \int dx \ \cos{\sqrt{\frac{4\pi}{K_{2}}}\Phi_{2}}\cos
           {\sqrt{\frac{4\pi}{K_{3}}}\Phi_{3}}.
\end{eqnarray}
Here, we neglect the gauge coupling temporarily and will discuss it later. The 
relevancy of these interactions is determined by their scaling dimensions. They 
are $\Delta_{1}=\frac{1}{K_{3}}+\frac{1}{K_{4}}$ and $\Delta_{2}=\frac{1}
{K_{2}}+\frac{1}{K_{3}}$, which correspond to $g^{'}_{1}$- and 
$g^{'}_{2}$-term, respectively. From eq.(\ref{al}), $K_{2} <1$ and $K_{4} >1$ 
at the isotropic point. This implies $\Delta_{1} <2$. Therefore, $g^{'}_{1}$-term is 
a relevant operator in the sense of renormalization group (RG). $\Phi_{3}$- and 
$\Phi_{4}$-field are pinned at some values and both acquire gaps. They are 
decoupled from the low energy theory. Taking into account these facts, the effective 
Hamiltonian becomes
\begin{equation}
 H_{h}=\frac{1}{2} \ \displaystyle{\sum_{\alpha =1,2}}K_{\alpha} \ \int dx \ [
       (\partial_{x}\Theta_{\alpha})^{2}+(\partial_{x}\Phi_{\alpha})^{2}]-g \ 
       \int dx \ \cos{\sqrt{\frac{4\pi}{K_{2}}}\Phi_{2}}
\end{equation}
where $g=2g^{'}_{2}\langle \cos{\sqrt{\frac{4\pi}{K_{3}}}\Phi_{3}}\rangle$. In 
the large spin limit, $K_{2} > 1/2$. Thus the scaling dimension of $g$-term is 
less than two. It is a relevant operator. The $\Phi_{2}$-field is also massive. 
In the low energy limit, we obtain our effective Hamiltonian as the following
\begin{equation}
 H_{eff}=\frac{K_{1}}{2} \ \int dx \ [(\partial_{x}\Theta_{1})^{2}+(\partial_{x}
         \Phi_{1})^{2}].
\end{equation}

  Now we would like to discuss the implications of our results. First of all, 
the low energy effective Hamiltonian consists of a gapless spin-singlet mode and 
from eq.(\ref{ak}), this mode describes the total charge density fluctuation. 
Furthermore, there is only one free parameter $K_{1}$, the compactification 
radius of $\Phi_{1}$-field, which has to be determined from experiments. This 
supports the suggestion proposed in ref.\cite{Troy}. Second, since the 
$\Phi_{3}$-field is pinned at some value, it can not affect the non-linear sigma 
model too much. Especially, it is unable to screen the long range Coulomb force. 
As a result, the gauge field is in the confining phase. This leads to two types of 
spin excitations. One is the massive triplet (In the $CP^{1}$ language, it is the
$z \bar{z}$ bound state), which is the same as the magnon in the undoped case. 
The other is the bound state of holes and $z$ quanta, which carries the same 
quantum numbers as the electron. In fact, this excitation can be considered as the 
breaking of a hole pair and we have two quasiparticles. Each carries electric 
charge one and spin one-half. The latter is also observed in ref.\cite{Troy}. In that 
paper, Troyer {\it et al.} find that the spin gap is determined by the bound state of 
spinons and holons. We cannot compare the gaps of these excitations in our 
approach. However, we give a picture about the formation of the unexpected spin 
excitation. The Berry phase term gives rise to the necessary attractive force between 
spinons and holons. Since this force is a gauge interaction and the latter in 1d is a 
linear confining potential, this results in bound states of spinons and holons. 

  After we understand the spectrum, we can calculate the asymptotic behavior of
various correlation functions. The most important ones are pairing and CDW 
correlation function. The definition of the pair-field is the same as eq.(\ref{ah}). 
The long distance behavior of the pairing correlation function can be calculated by the 
operator
\begin{equation}
 \Delta(j)\sim \exp{(-i\sqrt{\pi K_{1}}\Theta_{1})}
\end{equation}
because other fields are massive. The result is
\begin{equation}
 \langle \Delta(j)\Delta^{+}(0)\rangle \stackrel{\mid j \mid \rightarrow \infty}
 {\longrightarrow}\frac{1}{\mid j \mid^{K_{1}/2}}.
\label{am}
\end{equation}
Next we shall examine the $2k_{F}$ CDW susceptibility. The corresponding
order parameter $O_{CDW}$ can be expressed as the product of hole operators
and spin operators. As emphasized in the previous paragraph, matter fields do 
nothing much on the spin sector of the effective action. We can find the contributions
of spin operators from results of the undoped case. The work of Shelton {\it et al.}
\cite{SNT} showed that the low energy theory of two-leg ladders is described by 
four decoupled non-critical Ising models with three of them in the ordered phase
and one in the disordered phase (or vice versa). Moreover, the spin part in 
$O_{CDW}$ can be expressed in terms of the order ($\sigma$) and disorder 
($\mu$) parameter fields of the Ising model as 
$\mu_{1}\mu_{2}\mu_{3}\mu_{0}$ or 
$\sigma_{1}\sigma_{2}\sigma_{3}\sigma_{0}$. It is clear that the correlation
functions of the above operators decay exponentially to zero. This implies that 
$\langle O_{CDW}(x)O_{CDW}(0)\rangle$ shows the same behavior. Then we
have to consider the $4k_{F}$ CDW susceptibility or $\langle O_{CDW}^{2}(x)
O_{CDW}^{2}(0)\rangle$. Using the OPE $\mu(z)\mu(\omega)\sim 1/(z-\omega)
^{1/4}+ \cdots$, it is straightforward to see that the spin part of this correlation
function contributes a non-vanishing constant to it. Thus, the long distance behavior
of $\langle O_{CDW}^{2}(x)O_{CDW}^{2}(0)\rangle$ can be determined solely 
through its hole part. If we define the operator
\begin{eqnarray}
 O_{h}(x) &\equiv& \psi_{R,A,1}^{+}\psi_{R,B,2}^{+}\psi_{L,A,1}\psi_{L,B,2}
                                   \nonumber \\
                 &\sim& \exp{(-i\sqrt{\frac{4\pi}{K_{1}}}\Phi_{1})},
\label{an}
\end{eqnarray}
then 
\begin{eqnarray}
 \langle O_{CDW}^{2}(x)O_{CDW}^{2}(0)\rangle &\sim& \langle O_{h}(x)
   O_{h}(0)\rangle \nonumber \\
        &\sim& \frac{1}{\mid x \mid^{2/K_{1}}}.
\label{ao}
\end{eqnarray}
In the second line of eq.(\ref{an}), we keep the gapless mode only. From eqs.(\ref{am}) 
and (\ref{ao}), it is clear that SC dominates when $K_{1} < 2$ and CDW dominates 
when $K_{1} > 2$. In the large spin limit, $K_{1}\approx 1$. Therefore, we conclude 
that SC susceptibility dominates in two-leg $t-J$ ladders and a weak interladder 
interaction will lead to superconductivity at low temperature. Also the exponents of 
pairing and CDW susceptibility satisfy the relation: $K_{SC}\cdot K_{CDW}=1$. We 
arrive at the same conclusions as previous numerical investigations. Although we heavily 
rely on the large spin approximation, our results should capture the basic feature of the 
system with spin one-half.

    In summary, after we take $H_{1}$ and $H_{2}$ into account, the low energy
effective Hamiltonian only consists of one gapless charge mode, which describes 
the charge density fluctuation. The spin excitations are electron-like quasiparticles
and magnons and both have energy gaps. The $2k_{F}$ CDW susceptibility shows 
exponentially decaying behavior while those of  $4k_{F}$ CDW and singlet SC both 
show power-law behavior. With the above results, we conclude that lightly doped 
two-leg ladders fall into the universality class of Luther-Emery model. In addition,
this phase is dominated by singlet SC susceptibility according to our analysis.

  There are two related works which should be mentioned. The first is the paper by 
Ichinose and Matsui\cite{Ichi}. They also adopted the slave-fermion and $CP^{1}$ 
boson technique to treat the two-leg $t-J$ ladder. The other is the work of Ivanov 
and Lee\cite{Ivan}. They use the slave-boson scheme and arrive at the same 
conclusion that the low energy sector of the two-leg $t-J$ ladder is a Luther-Emery
liquid. For quantitative description of this system, they introduce a new order 
parameter $n_{pair}\equiv (n_{1}-\delta)(n_{2}-\delta)$ where $n_{i}$ is the hole 
density in the {\it i}-th chain in addition to the ones we considered. They suggested 
that the long distance behavior of the $2k_{F}$ part of $\langle n_{pair}(x)n_{pair}
(0)\rangle$ (Notice that their $2k_{F}$ is equivalent to our $4k_{F}$.) is as the 
following
\begin{eqnarray}
 \frac{A}{x^{\alpha_{1}}}+\frac{B}{x^{\alpha_{1}+2}} \nonumber
\end{eqnarray}
where A and B depend on the average pair overlap: $A \gg B$ at $\xi_{pair}
\delta \ll 1$ and $A \ll B$ at $\xi_{pair}\delta \gg 1$. Here $\xi_{pair}$ is the 
size of the hole pair and is of the order of the lattice spacing. $\alpha_{1}$ is the
exponent of $\langle O_{CDW}^{2}(x)O_{CDW}^{2}(0)\rangle$. We calculate this 
correlation function and we obtain a power-law behavior with the exponent equal to 
$2/K_{1}$, which is the same as the one of 
$\langle O_{CDW}^{2}(x)O_{CDW}^{2}(0)\rangle$. We do not find an exponent
equal to $2/K_{1}+2$ even including higher harmonics. Since our approach is valid 
when $\delta \ll 1$, we cannot tell whether the second term exists or not at larger 
doping concentration. 

\section{Conclusion}
\vskip 0.5cm

  Anderson has proposed that the spin liquid state may evolve into a superconductor 
upon doping. However, it has been proved notoriously difficult to have any 
concrete analytic result to confirm this idea in 2D. A doped spin ladder may 
provide a good place to study this mechanism though there is no true long range 
order here. This is so because the undoped two-leg ladder is a kind of spin liquid state 
and it is simpler to deal with this problem both analytically and numerically. Soon
after the discovery of high $T_{c}$ superconductivity a model was proposed to 
describe the doped spin liquid state by coupling holes to the non-linear sigma model 
and the gauge interaction between A and B holes provides the necessary attraction for 
pairing. Our approach is basically to apply the same idea to ladder systems. In contrast 
to previous studies, the gauge interaction plays a minor role on the formation of hole 
pairing in two-leg $t-J$ ladders. Instead, it is mainly due to the effective attraction 
between the nearest-neighbor holes. Therefore, upon doping, these holes are inclined 
to stay on the same preexisting singlets. Although this may be a unique characteristic 
of two-leg ladders exclusively, this point deserves further examination in higher 
dimensions. We also point out that the gauge interaction is responsible for the 
electron-like collective mode found in ref.\cite{Troy}. The existence of this type of 
excitations is independent of the quartic fermion interactions. The absence or presence 
of the latter determines whether the long range Coulomb force is screened or not and 
thus affects the magnitude of the gap of excitations with the non-trivial spin quantum 
number.  More importantly, it does affect the existence of the magnon: there are no 
magnons without the quartic fermion interactions. We have to emphasize that both the 
effective attraction between nearest-neighbor holes and gauge interactions are due to 
the strong short-range AF background, which has its origin in the strong repulsive 
interactions between electrons. 

  The main assumption we made is to replace eq.(\ref{ac}) with eqs.(\ref{ae}) and 
(\ref{af}). This is justified for undoped case and for lightly doped two-leg ladders. The 
spin excitations in the undoped three-leg ladder are gapless and thus the distortion of 
the short-range AF background arising from hole motions will be so serious such that 
eq.(\ref{ad}) may be far from the real situation. Moreover, there may be no coherent 
peak in the spectral function of one-particle Green function and we are unable to 
simply add a $t'$-term to the effective Hamiltonian. In other words, the $t$-term should 
play a more important role in this case. As has been shown in ref.\cite{Rice2}, upon 
doping, the three-leg $t-J$ ladder has two components - - a conducting Luttinger liquid 
coexisting with an insulating spin liquid phase. In order to discuss its low energy 
physics in the same spirit as the present paper, we need more understanding about the 
ground state of the undoped case and the behavior of holes in this ground state. Work 
along this line is under progress.

\appendix
\section{Bosonization Rules}
\vskip 0.5cm

    We list all bosonization rules we used in the following:
\begin{eqnarray}
 \psi_{L}(z) &=& \frac{1}{\sqrt{2\pi a}}exp\{-i\sqrt{4\pi}\phi_{L}(z)\}, \nonumber
                            \\
 \psi_{R}(\bar{z}) &=& \frac{1}{\sqrt{2\pi a}}exp\{i\sqrt{4\pi}\phi_{R}(\bar{z})\}, 
                                      \nonumber \\
  i\bar{\psi}\gamma^{\mu}\partial_{\mu}\psi &=& \frac{1}{2}(\partial_{\mu}\phi)^{2}, 
                                      \nonumber \\
  :\bar{\psi}\gamma_{\mu}\psi: &=& \frac{1}{\sqrt{\pi}}\epsilon_{\mu \nu}\partial^{\nu}\phi 
                                      \nonumber
\end{eqnarray}
where $z=\tau +ix$, $\bar{z}$ is its complex conjugate and $a$ is the short distance cut-off.
$\psi_{L}(z)$ and $\psi_{R}(\bar{z})$ are left and right fermions, respectively. $\phi_{L}(z)$
and $\phi_{R}(\bar{z})$ are bosonic left and right movers, respectively. In terms of them, we
define $\phi(\tau ,x)=\phi_{L}+\phi_{R}$ and $\Theta(\tau ,x)=\phi_{L}-\phi_{R}$. 
\vskip 0.5cm

\begin{center}
 {\bf Acknowledgements}
\end{center}
\vskip 0.5cm

 We would like to acknowledge useful discussions with X.G.Wen and Y.C.Kao. We also wish 
to thank the Center for Theoretical Science of National Science Council of R.O.C. for partial 
supports. This work is supported by National Science Council of R.O.C. under grant No. 
NSC87-2811-M-007-0045.


\begin{thebibliography}{99}
\bibitem{Higg} In the use of these terminology, we follow the definition of the paper:
                         D.J.Gross, I.R.Klebanov, A.V.Matytsin and A.V.Smilga, Nucl.Phys.
                         {\bf B461}, 109 (1996). For the $1+1$ dimensional gauge theories, 
                         the Higgs phase means that the potential $V(R)$ between two 
                         fractional probe charges separated by a large distance $R$ is of the 
                         form $V(R)\sim e^{-\mu R}/R$, $\mu$ being the screening mass. On 
                         the other hand, we say that the gauge fields are in the confining phase 
                         when $V(R)\sim \sigma R$ with a non-vanishing $\sigma$.
\end{thebibliography}
\end{document}